\documentclass[aps,pra,twocolumn,showpacs]{revtex4-1}

\usepackage{bm,graphicx,textcomp,amssymb,amsmath,dcolumn,color}

\newcommand{\ket}[1]{\left|#1\right>}
\newcommand{\bra}[1]{\left<#1\right|}

\newcommand{\nn}{\nonumber\\}

\newcommand{\f}[1]{\mbox{\boldmath$#1$}}

\newcommand{\bea}{\begin{eqnarray}}
\newcommand{\ea}{\end{eqnarray}}
\newcommand{\eea}{\end{eqnarray}}
\newcommand{\ord}{\,{\cal O}}

\begin{document}

\title{Multiple particle-hole pair creation in the Fermi-Hubbard model
by a pump laser}

\author{Nicolai ten Brinke, Manuel Ligges, Uwe Bovensiepen and
Ralf~Sch\"utzhold}

\affiliation{Fakult\"at f\"ur Physik,
Universit\"at Duisburg-Essen, Lotharstra{\ss}e 1, Duisburg 47048, Germany}

\date{\today}

\begin{abstract}
We study the Fermi-Hubbard model in the strongly correlated Mott phase under
the influence of a harmonically oscillating electric field, e.g., a pump laser.
In the Peierls representation, this pump field can be represented as an
oscillating phase of the hopping rate $J(t)$, such that the effective
time-averaged rate $\bar J$ is reduced, i.e., switching the pump laser
suddenly is analogous to a quantum quench.
Apart from this time-averaged rate $\bar J$, it is well known that the
oscillating component of $J(t)$ can resonantly create particle-hole pairs if
the pump frequency $\omega_{\rm pump}$ equals (or a little exceeds) the Mott gap.
In addition, we find that it is possible to create multiple pairs if
$\omega_{\rm pump}$ is near an integer multiple of the gap.
These findings should be relevant for pump-probe experiments.
\end{abstract}


\maketitle

\section{Introduction}

Understanding the non-equilibrium dynamics of strongly correlated quantum
many-body systems is a major challenge in physics -- both from the theoretical
and the experimental point of view.
As a prototypical example, we shall consider the Fermi-Hubbard model
\cite{Hubbard}
\bea
\label{Fermi-Hubbard}
\hat H_{\rm FH}
=
-J\sum\limits_{<\mu,\nu>,s}\hat c^\dagger_{\mu,s}\hat c_{\nu,s}
+U\sum\limits_\mu\hat n_\mu^\uparrow\hat n_\mu^\downarrow
\,.
\ea
Here $\hat c^\dagger_{\mu,s}$ and $\hat c_{\nu,s}$ are the fermionic creation
and annihilation operators at the neighboring lattice sites $\mu$ and $\nu$
with the spin $s$ while $\hat n_\mu^s$ is the corresponding number operator.
We consider half filling
$\langle\hat n_\mu^\uparrow\rangle=\langle\hat n_\mu^\downarrow\rangle=1/2$
and assume that the hopping rate $J$ is much smaller than the on-site
repulsion $U$ which ensures that we are in the Mott insulator phase
with a Mott gap of roughly $U$.



A nice and conceptually clear example for a non-equilibrium situation is a
quantum quench where the system starts in the ground (or thermal equilibrium)
state and one of the parameters such as the hopping rate $J$ is changed
suddenly.
After that, the system is no longer in its ground or equilibrium state in
general and the subsequent dynamics including phenomena like relaxation,
pre-thermalization and thermalization have been studied in various works,
see, e.g.,
\cite{Moeckel+Kehrein-08,Moeckel+Kehrein-09,Eckstein+Kollar+Werner-09,
Eckstein+Kollar+Werner-10,Schiro+Fabrizio,Propagation,Equilibration}.
In the following, we shall consider a somewhat different kind of
non-equilibrium dynamics which is induced by an oscillating field and thus
gives rise to oscillating parameters such as $J(t)$, see also
\cite{Tokuno+Demler+Giamarchi,Frank}, for example.
However, as we shall find below, some aspects are similar to quantum
quenches.

\section{Pump beam}

For simplicity, let us start with the Fermi-Hubbard Hamiltonian in one
dimension ($\hbar=1$)
\bea
\label{Fermi-Hubbard-1D}
\hat H_{\rm FH}^{\rm 1D}
=
-J\sum\limits_{\mu,s}
\left(
\hat c^\dagger_{\mu,s}\hat c_{\mu+1,s}
+{\rm h.c.}
\right)
+U\sum\limits_\mu\hat n_\mu^\uparrow\hat n_\mu^\downarrow
\,.
\ea
Neglecting the magnetic component of the pump laser, the most obvious impact
of the pump field (others will be discussed below)
is a time-dependent shift of the energies corresponding to the Hamiltonian
\bea
\label{pump-impact}
\hat H_{\rm pump}(t)
=
\sum\limits_\mu
\left(\hat n_\mu^\uparrow+\hat n_\mu^\downarrow \right)
V_\mu(t)
\,.
\ea
Hamiltonians of this kind are often discussed in the context of driven
quantum lattice systems.
Assuming that the laser wavenumber $k_{\rm laser}^\|$ parallel to the lattice
is small compared to the other relevant scales, the site-dependent energy
shift $V_\mu(t)\approx-q\f{r}_\mu\cdot\f{E}_{\rm pump}(t)$
at the position $\f{r}_\mu$ of the site $\mu$
is determined by the electric pump field $\f{E}_{\rm pump}(t)$
with $q$ being the elementary charge.
This Hamiltonian~\eqref{pump-impact} generates the Peierls 
transformation
\bea
\hat c_{\mu,s}(t)\to\hat c_{\mu,s}(t)e^{i\varphi_\mu(t)}
\,,
\ea
with the time-dependent phase $\dot\varphi_\mu(t)=V_\mu(t)$.
Inserting this transformation back into Eq.~\eqref{Fermi-Hubbard},
we find that the tunneling term $\propto J$ acquires an oscillating phase
\bea
\label{oscillating-phase}
J\to J(t)=J_0 e^{i\Delta\varphi(t)}
\,.
\ea
Assuming a harmonically oscillating time-dependence, we may insert
$E_{\rm pump}^\|(t)=E_{\rm pump}^\|\cos(\omega_{\rm pump}t)$ and obtain
\bea
\label{deltavarphi}
\Delta\varphi(t)
=
q\ell E_{\rm pump}^\|
\frac{\sin(\omega_{\rm pump}t)}{\omega_{\rm pump}}
=
\Delta\varphi_{\rm max}\sin(\omega_{\rm pump}t)
\,,
\ea
with the lattice spacing $\ell$.

\subsection{Effective quantum quench}

If the pump frequency $\omega_{\rm pump}$ is much larger than all the other
relevant energy scales such as $J$ and $U$, the main consequence of the
time-dependence~\eqref{oscillating-phase} is that the original hopping
rate $J$ in the Hamiltonian~\eqref{Fermi-Hubbard-1D} can effectively be
replaced by the time-averaged hopping rate $\bar J$.
For a harmonic oscillation, we may calculate the time average via the
Jacobi-Auger expansion and obtain
\bea
\bar J
=
\overline{J_0 e^{i\Delta\varphi(t)}}
=
J_0 {\mathfrak J}_0(\Delta\varphi_{\rm max})
\,,
\ea
where ${\mathfrak J}_0$ denotes the Bessel function of the first kind.
Since $|{\mathfrak J}_0|\leq1$, the effective time-averaged hopping rate
is lowered by the pump beam.
For certain values of $\Delta\varphi_{\rm max}$ such as
$\Delta\varphi_{\rm max}^0\approx2.4$, one may even effectively inhibit hopping
due to ${\mathfrak J}_0(\Delta\varphi_{\rm max}^0)=0$.
Thus, if we would switch on (or off) the pump beam sufficiently fast --
i.e., faster that the characteristic response time of our system --
the situation would be very analogous to a quantum quench as discussed in
\cite{Moeckel+Kehrein-08,Moeckel+Kehrein-09,Eckstein+Kollar+Werner-09,
Eckstein+Kollar+Werner-10,Schiro+Fabrizio,Propagation,Equilibration},
for example.
As shown in these papers, such a quench will create particle-hole
(doublon-holon) pairs in general -- the number (density) of those pairs
will depend on the parameters such as $U$ and the initial $J_{\rm in}$
and final $J_{\rm out}$ hopping rates.

If the phase $\Delta\varphi_{\rm max}$ is small, a Taylor expansion gives
\bea
\bar J\approx J_0\left(1-\frac12\,\overline{\Delta\varphi^2(t)}\right)
=
J_0\left(1-\frac14\,\Delta\varphi^2_{\rm max}\right)
\,.
\ea
In this case, the change of the hopping rate is relatively small
$\Delta\bar J=-J_0\Delta\varphi^2_{\rm max}/4$
and thus we may employ time-dependent perturbation theory where the
perturbation Hamiltonian is governed by $\Delta\bar J$.
As the perturbation Hamiltonian scales quadratically in $\Delta\varphi\ll1$
and thus linearly in the pump intensity $I_{\rm pump}\propto E_{\rm pump}^2$,
the probability for pair creation (per unit length) would be suppressed
as the fourth power of $\Delta\varphi\ll1$, i.e., it would scale
quadratically in the pump intensity $P\propto E_{\rm pump}^4 \propto I_{\rm pump}^2$.
This scaling could help to distinguish the above quench mechanism from other
effects (which scale linearly in $I_{\rm pump}$, for example).

\subsection{Wave-functions}

One should also keep in mind that the Hamiltonian~\eqref{pump-impact} only
contains the component of the electric field parallel to the lattice --
while the perpendicular component can also induce effects such as the
deformation of wave-functions leading to variations of $J$ and $U$,
i.e., it can also cause small oscillations in $J$ and $U$.
However, assuming that the initial state is the ground state
(i.e., an eigenstate) of the Hamiltonian~\eqref{Fermi-Hubbard-1D},
the perturbation caused by a small variation of $U(t)$ is equivalent
(to lowest order)
to the perturbation caused by an appropriate small variation of $J(t)$.
Thus, in the following, we shall consider the following general
perturbation Hamiltonian
\bea
\label{perturbation}
\hat H_{\Delta J}(t)
=
-\sum\limits_{\mu,s}
\left(
\Delta J(t)
\hat c^\dagger_{\mu,s}\hat c_{\mu+1,s}
+{\rm h.c.}
\right)
\,,
\ea
where $\Delta J(t)$ could be a real or complex oscillating function.
In higher-dimensional lattices, $\Delta J(t)$ can also depend on the lattice
indices $\Delta J_{\mu\nu}(t)$, e.g., on the direction relative to the
pump beam, but we shall omit this dependence for simplicity here.
More generally, repeating the steps of the derivation of the Fermi-Hubbard
Hamiltonian~\eqref{Fermi-Hubbard} from the underlying many-body Hamiltonian
(including the Coulomb interaction) in the presence of the pump field,
one would also obtain oscillating terms like
$\hat c^\dagger_{\mu,s}\hat c^\dagger_{\nu,s'}W_{\mu\nu\lambda\sigma}^{ss'}(t)
\hat c_{\lambda,s}\hat c_{\sigma,s'}$, but we shall also not consider these
contributions here.

\section{Single pair creation}

Apart from the reduction of the time-averaged $\bar J$, the oscillating
contribution
$\Delta J(t)\propto\Delta\varphi(t)\propto\sin(\omega_{\rm pump}t)$
can also have an impact.
For example, if $\omega_{\rm pump}$ satisfies the resonance condition
$\omega_{\rm pump}=U$, it would resonantly create particle-hole pairs.
As one way to understand this process, let us return to the more general
Fermi-Hubbard model~(\ref{Fermi-Hubbard}) and employ the hierarchy of
correlations discussed in \cite{Emergence,Correlations,Quasi-particle}.
To this end, we consider the reduced density matrix $\hat\varrho_\mu$
of one lattice site $\mu$ and analogously $\hat\varrho_{\mu\nu}$ for two
lattice sites $\mu$ and $\nu$ etc.
Separating the correlated part via
$\hat\varrho_{\mu\nu}=
\hat\varrho_{\mu\nu}^{\rm corr}+\hat\varrho_\mu\hat\varrho_\nu$,
we may derive the evolution equations for
$\partial_t\hat\varrho_{\mu\nu}^{\rm corr}$ etc.
To lowest order, the ground state (Mott insulator) restricted to two lattices
sites can be represented by the equipartition state
$\ket{\uparrow,\downarrow}_{\mu\nu}=
\hat c^\dagger_{\mu,\uparrow}\hat c^\dagger_{\nu,\downarrow}\ket{0}$
while the state with a doublon-holon excitation at these two sites
can be written as
$\ket{\uparrow\!\downarrow,0}_{\mu\nu}=
\hat c^\dagger_{\mu,\uparrow}\hat c^\dagger_{\mu,\downarrow}\ket{0}$.
Calculating the matrix element of $\hat\varrho_{\mu\nu}^{\rm corr}(t)$
between these two states, we find
\bea
\bra{\uparrow\!\downarrow,0}
(i\partial_t-U)\hat\varrho_{\mu\nu}^{\rm corr}(t)
\ket{\uparrow,\downarrow}=J(t){\cal M}_{\mu\nu}^{(2)}
\,,
\ea
where ${\cal M}_{\mu\nu}^{(2)}$ denotes a matrix element containing
the on-site matrices
$\hat\varrho_\mu$ and $\hat\varrho_\nu$, for example,
cf.~\cite{Emergence,Correlations,Quasi-particle}.
Again, $\Delta J(t)$ can also depend on the lattice indices
$\Delta J_{\mu\nu}(t)$, but we shall omit this here.
Evidently, if $J(t)=J_0+\Delta J(t)$ [or $\Delta J_{\mu\nu}(t)$]
oscillates with the frequency $\omega_{\rm pump}=U$,
we would get a resonant growth of $\hat\varrho_{\mu\nu}^{\rm corr}(t)$
corresponding to particle-hole (doublon-holon) pair creation.

\section{Double pair creation}

This well-known resonance condition $\omega_{\rm pump}=U$ is not the only
possibility.
As we shall demonstrate below, for $\omega_{\rm pump}=2U$, one could resonantly
create two particle-hole pairs at the same time, for example.
This effect can be understood analogously in terms of the four-point
correlator $\hat\varrho_{\mu\nu\lambda\sigma}^{\rm corr}$ whose matrix element
obeys the equation
\bea
\bra{\uparrow\!\downarrow,0,\uparrow\!\downarrow,0}
(i\partial_t-2U)\hat\varrho_{\mu\nu\lambda\sigma}^{\rm corr}(t)
\ket{\uparrow,\downarrow,\uparrow,\downarrow}
=J(t){\cal M}_{\mu\nu\lambda\sigma}^{(4)}
\,.
\ea
The remaining matrix element ${\cal M}_{\mu\nu\lambda\sigma}^{(4)}$ contains
products of two-point correlations such as
$\hat\varrho_{\mu\nu}^{\rm corr}\hat\varrho_{\lambda\sigma}^{\rm corr}$.
Thus, we necessarily obtain resonant creation of two particle-hole
(doublon-holon) pairs at the same time -- unless the
source term ${\cal M}_{\mu\nu\lambda\sigma}^{(4)}$ vanishes identically.

In order to show that this source term is non-vanishing, let us consider a
simple and exactly solvable case -- the Fermi-Hubbard
model~\eqref{Fermi-Hubbard} on a tetrahedron, i.e., two spin-up plus two
spin-down fermions on four lattice sites with full permutation invariance.
For vanishing hopping $J=0$, the ground state is the fully symmetrized state
$\ket{\psi_0}=\ket{\uparrow,\downarrow,\uparrow,\downarrow}_{\rm symm}$.
Analogously, the first excited state reads
$\ket{\psi_1}=\ket{\uparrow\!\downarrow,0,\uparrow,\downarrow}_{\rm symm}$
and the highest energy state is
$\ket{\psi_2}=\ket{\uparrow\!\downarrow,0,\uparrow\!\downarrow,0}_{\rm symm}$.
In this case $J=0$, the matrix element
$\bra{\psi_2}\hat H_{\Delta J}\ket{\psi_0}$
would be zero since one cannot go from
$\ket{\psi_0}$ to $\ket{\psi_2}$
with only one hopping event.

For small $J>0$, however, the ground state also contains a small $\ord(J)$
admixture of $\ket{\psi_1}$ and an even smaller $\ord(J^2)$ of $\ket{\psi_2}$.
As one way to see this, one can exactly diagonalize the
Hamiltonian~\eqref{Fermi-Hubbard} for this simple case.
Using the three vectors $\ket{\psi_0}$, $\ket{\psi_1}$, and $\ket{\psi_2}$
as a basis for the fully permutation-invariant sub-space of the Hilbert space,
the Hamiltonian~\eqref{Fermi-Hubbard} can be represented by a
$3\times3$-matrix of the following form
\bea
\hat H_{\rm FH}
=
\begin{pmatrix}
0 & -4J & 0 \\
-4J & U -4J & -4J \\
0 & -4J & 2 U
\end{pmatrix}
\,.
\ea
Diagonalization of this matrix yields the ground state
(for small but non-zero values of $J$)
\bea
\ket{\psi}_{\rm ground}
&=&
\left(1-\frac{J^2}{2U^2}\right)\ket{\psi_0}
+
\left(\frac{J}{U}+\frac{J^2}{U^2}\right)\ket{\psi_1}
\nn
&&
+
\frac{J^2}{2U^2}\ket{\psi_2}
+
\ord\left(\frac{J^3}{U^3}\right)
\,.
\ea
E.g., if we suddenly switched off $J$ (quantum quench), this admixture of
$\ket{\psi_1}$ or $\ket{\psi_2}$ contained in $\ket{\psi}_{\rm ground}$
would then yield the amplitude for creating one or two pairs by this
quantum quench.
Analogous expressions can be derived for the first excited state
$\ket{\psi}_{\rm first}$ containing one particle-hole pair and
the highest-energy state $\ket{\psi}_{\rm highest}$ containing
two particle-hole pairs.
Now, calculating the matrix element of the perturbation
Hamiltonian~\eqref{perturbation} $\propto\Delta J(t)$ between the ground
state and the highest energy sate --
which corresponds to the resonant generation of two pairs at the same time
-- we find that these admixtures yield a non-zero amplitude
\bea
\bra{\psi_{\rm highest}}\hat H_{\Delta J}\ket{\psi_{\rm ground}}
=
\ord\left(\Delta J\,\frac{J^2}{U^2}\right)
\,.
\ea
Of course, this simple model does not model a realistic lattice in a
solid-state setting, for example, but it shows that the source term
${\cal M}_{\mu\nu\lambda\sigma}^{(4)}$ is non-zero, i.e.,
that one can create a double particle-hole pair with $\omega_{\rm pump}=2U$.
If the perturbation $\Delta J_{\mu\nu}(t)$ would depend on the lattice sites,
the associated Hamiltonian $\hat H_{\Delta J}$ would not preserve the full
permutational invariance in general.
However, the above results would the still apply to the projection of
$\hat H_{\Delta J}$ to the fully permutation-invariant sub-space of the
Hilbert space -- which is sufficient to prove a non-zero probability.

This double pair creation phenomenon is enabled by the interplay of hopping
$J$ and interaction $U$ or, alternatively, of the correlation between sites
(due to $J$) and the correlation between particles (due to $U$).
Consistently, this effect vanishes both for $J=0$ and for $U=0$ and has
maximum probability for intermediate values of $J/U\approx0.3$.
Thus, such a signal would be a signature of quantum correlations.
Note, that, in contrast to two-photon or multi-photon effects (Floquet theory)
with the resonance condition $2\omega_{\rm pump}=U$ or $n\omega_{\rm pump}=U$,
this is a quantum effect more similar to parametric down-conversion in quantum
optics.

Interestingly, for the Fermi-Hubbard model on a square
(instead of a tetrahedron), we do not find this double pair-creation effect
-- at least not in the fully symmetric sub-space.
Whether this is a result of these symmetries, the reduced coordination number
(two instead of three), or the bi-partite structure of the square which
facilitates anti-ferromagnetic N\'eel ordering of the spins,
should be clarified in future investigations.
%

\section{Multiple pair creation}

As one might already have guessed, it is also possible to create three, four
or even more pairs for $\omega_{\rm pump}=3U$ $\omega_{\rm pump}=4U$ etc.
However, as these processes involve higher-order correlations --
e.g., for three pairs, one would have to consider the six-point correlator
-- they are more and more suppressed.
In analogy to the tetrahedron, we considered fully permutationally invariant
lattices with six and eight sites containing the same number of particles.
Again restricting ourselves to the fully permutationally invariant sub-space
of the Hilbert space, the Hamiltonian for six sites reads
\bea
\label{FH-6}
\hat H_{\rm FH}
=
\begin{pmatrix}
0 & -6J & 0 & 0 \\
-6 J & U -8J & -8J & 0 \\
0 & -8J & 2 U -8J & -6J  \\
0 & 0 & -6J & 3 U
\end{pmatrix}
\,,
\ea
and similarly for eight sites
\bea
\label{FH-8}
\hat H_{\rm FH}
=
\begin{pmatrix}
0 & -8 {J} & 0 & 0 & 0 \\
-8 {J} & U -12 {J} & -12 {J} & 0 & 0 \\
0 & -12 {J} & 2 U -16 {J} & -12 {J} & 0  \\
0 & 0 & -12 {J} & 3 U -12 {J} & -8 {J} \\
0 & 0 & 0 & -8 {J} & 4 U
\end{pmatrix}
\,.
\ea
For small $J$, we found that the three-pair amplitudes scale with
$\Delta J(J/U)^4$ for both Hamiltonians~\eqref{FH-6} and \eqref{FH-8},
while the four-particle amplitude behaves as $\Delta J(J/U)^6$
[for the Hamiltonian~\eqref{FH-8}].
Again, the matrix elements vanish for $J=0$ and $U=0$ and display a single
maximum at intermediate values of $J/U$.
These values of $J/U$ where the probabilities are maximal decrease with
increasing coordination number $Z$.

\section{Experimental realization}

We now consider a possible experimental realization of multiple particle-hole
pair creation as well as its spectroscopic evidence, based on femtosecond
time- and angle-resolved photo-emission spectroscopy (trARPES).
In trARPES, the sample under investigation is first excited using a rather
intense femtosecond optical pulse with central frequency $\omega_{\rm pump}$
and pulse duration $t_{\rm pump}$.
The generated non-equilibrium state is subsequently probed by means of direct
photo-emission using a second (weak) laser pulse
($\omega_{\rm probe}$, $t_{\rm probe}$).
The overall spectral and temporal  experimental resolution is then given by
the convolution of both pulses properties and limited by the time-band width
product $\Delta\omega\Delta t\geq 4\ln(2)$, resulting in typical values of
$\Delta t\approx$~50-150~fs and $\hbar\Delta\omega\approx$~20-50~meV
(Gaussian full width at half maximum).
In the strongly correlated Mott regime, these conditions allow for a
spectroscopic separation of the ground- and excited state signatures
(separated by the gap energy of approximately $U$ of typically a few hundred
meV) but fail to temporally resolve the full dynamics of individual or
multiple particle hole pairs that is expected to occur on time scales as
short as $\hbar/J\approx$~1~fs \cite{Eckstein08}.
Nevertheless, tracking the full dynamics is not a necessary prerequisite
for the effects under discussion here and it would be sufficient to observe
a temporally averaged signal in the corresponding energy window.

A prototypical Mott-insulator system that has being widely investigated
using trARPES (however, so far not under the conditions proposed here)
is the layered transition metal dichalcogenide 1\emph{T}-TaS$_2$
\cite{Perfetti06, Perfetti08, Petersen11, Hellmann12}.
The Mott transition in this system goes along with the formation of
commensurate charge density wave order and a periodic lattice distortion
\cite{Fazekas79}, leading to a superstructure formation with rather large
lattice spacing of $\ell$=1.23~nm in a hexagonal lattice ($Z=6$)
\cite{Yamamoto83}.
%
Assuming the on-site Coulomb repulsion $U=0.4~\rm eV$ and typical excitation
conditions ($E_{\rm pump}\approx1.4\cdot10^{8}~\rm V/m$) reported in
Refs.~\cite{Perfetti06, Perfetti08}, $J$ can be quenched efficiently.
Furthermore, the reported ratio $J/U\approx 0.7$ \cite{Perfetti08} is
favorable for multiple pair generation since neither $J$ nor $U$
are very small.
For example, in the fundamental resonance case $\omega_{\rm pump}=U$, 
the oscillating phase 
can be as large as $\Delta\varphi_{\rm max}\approx0.43$, and, correspondingly, 
$J$ can be transiently reduced by almost a factor of two. 
The effective (time averaged) tunneling rate $\bar J$ would then be 
decreased by more than four percent in comparison to $J_0$.
In the double-pair creation scenario, $\omega_{\rm pump}=2U$, $J$ would also be 
quenched in a non-negligible way ($\Delta\varphi_{\rm max}\approx0.22$), 
which results in an averaged tunneling rate $\bar J$ 
reduced by roughly one percent.

A reasonable experimental approach to verify the generation of multiple
particle-hole pairs from the absorption of single photons would be to verify
the systematic appearance and disappearance of the excited states
signature upon changes of the resonant pumping conditions.

\section{Conclusions}

We studied the influence of a pump laser on the Fermi-Hubbard model in the
strongly correlated Mott phase.
In the Peierls representation, the most obvious effect is an oscillating
hopping rate, which gives rise to two major effects:
First, due to a reduction of the effective time-averaged hopping rate,
switching on the pump laser is analogous to a quantum quench.
Second, the remaining oscillating contribution of $J(t)$ can resonantly
create particle-hole pairs.
In addition to the well-known fundamental resonance $\omega_{\rm pump}=U$,
we find higher resonances at $\omega_{\rm pump}=2U$ and $\omega_{\rm pump}=3U$
and so on, which correspond to the creation of multiple particle-hole pairs.

This multiple pair creation effect is caused by the interplay between the
correlations between particles (due to $U$) on the one hand and the
correlations between lattice sites (due to $J$) on the other hand.
Thus it is a genuine signature of these non-trivial correlations.
We also discussed experimental parameters which show that this effect should
be observable in pump-probe spectroscopy.
In this context, it is often implicitly assumed that efficient pumping is
only possible at the fundamental resonance $\omega_{\rm pump}=U$, while our
prediction suggests that this paradigm should be reconsidered.

Note that the multiple pair creation effect considered here is different
from charge carrier multiplication such as impact ionization, see, e.g.,
\cite{Werner+Held+Eckstein}, where an excitation generates further
particle-hole pairs {\em after} its creation.
In contrast, the effect considered here describes the generation of
multiple pairs {\em simultaneously} by one and the same photon.
In summary, we find that the non-equilibrium dynamics of strongly correlated
quantum many-body systems is still not fully understood and can afford
surprises.
This motivates further studies and development, e.g., regarding the theory
of pump-probe spectroscopy, see, e.g., \cite{Freericks+Krishnamurthy+Pruschke,
Freericks+Krishnamurthy+Sentef+Devereaux}.

\acknowledgments

R.S.\ and N.t.B.\ were supported by DFG (SFB-TR12)
and acknowledge fruitful discussions with K.~Krutitsky.




\begin{thebibliography}{99}

\bibitem{Hubbard}
J.~Hubbard, Proc.~R.~Soc.~Lond.~A {\bf 276}, 238 (1963);
ibid.\
{\bf 277}, 237 (1964);
ibid.\
{\bf 281}, 401 (1964).


\bibitem{Moeckel+Kehrein-08}
Michael Moeckel and Stefan Kehrein,
Phys.\ Rev.\ Lett.\ {\bf 100}, 175702 (2008).

\bibitem{Moeckel+Kehrein-09}
Michael Moeckel and Stefan Kehrein,
Annals of Physics {\bf 324}, 2146 (2009).

\bibitem{Eckstein+Kollar+Werner-09}
Martin Eckstein, Marcus Kollar, and Philipp Werner,
Phys.\ Rev.\ Lett.\ {\bf 103}, 056403 (2009).

\bibitem{Eckstein+Kollar+Werner-10}
Martin Eckstein, Marcus Kollar, and Philipp Werner,
Phys.\ Rev.\ B {\bf 81}, 115131 (2010).

\bibitem{Schiro+Fabrizio}
Marco Schir\'o and Michele Fabrizio,
Phys.\ Rev.\ Lett.\ {\bf 105}, 076401 (2010).

\bibitem{Propagation}
K.V.~Krutitsky, P.~Navez, F.~Queisser, R.~Sch\"utzhold,
EPJ Quantum Technology {\bf 1}, 12 (2014).

\bibitem{Equilibration}
F.~Queisser, K.V.~Krutitsky, P.~Navez, R.~Sch\"utzhold,
Phys.\ Rev.\ A {\bf 89}, 033616 (2014).


\bibitem{Tokuno+Demler+Giamarchi}
Akiyuki Tokuno, Eugene Demler, and Thierry Giamarchi,
Phys.\ Rev.\ A {\bf 85}, 053601 (2012).

\bibitem{Frank}
Regine Frank,
New J.\ Phys.\ {\bf 15}, 123030 (2013).


\bibitem{Emergence}
P.~Navez, R.~Sch\"utzhold,
Phys.\ Rev.\ A {\bf 82}, 063603 (2010).

\bibitem{Correlations}
F.~Queisser, K.V.~Krutitsky, P.~Navez, R.~Sch\"utzhold,
{\tt arXiv:1203.2164}.

\bibitem{Quasi-particle}
P.~Navez, F.~Queisser, R.~Sch\"utzhold,
J.\ Phys.\ A {\bf 47}, 225004 (2014).


\bibitem{Eckstein08}
M. Eckstein, M. Kollar,
Phys.\ Rev.\ B {\bf 78}, 245113 (2008).

\bibitem{Perfetti06}
L.~Perfetti, P.A.~Loukakos, M.~Lisowski, U.~Bovensiepen, H.~Berger,
S.~Biermann, P.S.~Cornaglia, A.~Georges, M.~Wolf,
Phys.\ Rev.\ Lett. {\bf 97}, 067402 (2006).

\bibitem{Perfetti08}
L.~Perfetti, P.A.~Loukakos, M.~Lisowski, U.~Bovensiepen, M.~Wolf,
H.~Berger, S.~Biermann, A.~Georges,
New\ J.\ Phys. {\bf 10}, 053019 (2008).

\bibitem{Hellmann12}
S.~Hellmann, T.~Rohwer, M.~Kall\"{a}ne, K.~Hanff, C.~Sohrt, A.~Stange,
A.~Carr, M.M.~Murnane, H.C.~Kapteyn, L.~Kipp, M.~Bauer, K.~Rossnagel,
Nat.\ Commun. {\bf 3}, 1069 (2012).

\bibitem{Petersen11}
J.C.~Petersen, S.~Kaiser, N.~Dean, A.~Simoncig, H.Y.~Liu,
A.L.~Cavalieri, C.~Cacho, I.C.E.~Turcu, E.~Springate, F.~Frassetto,
L.~Poletto, S.S.~Dhesi, H.~Berger, A.~Cavalleri,
Phys.\ Rev.\ Lett. {\bf 107}, 177402 (2011).

\bibitem{Fazekas79}
P.~Fazekas, E.~Tosatti,
Philos.\ Mag.\ B {\bf 39}, 229 (1979).

\bibitem{Yamamoto83}
A.~Yamamoto,
Phys.\ Rev.\ B {\bf 27}, 7823 (1983).


\bibitem{Werner+Held+Eckstein}
Philipp Werner, Karsten Held, and Martin Eckstein,
Phys.\ Rev.\ B {\bf 90}, 235102 (2014).


\bibitem{Freericks+Krishnamurthy+Pruschke}
J.K.~Freericks, H.R.~Krishnamurthy, and Th.~Pruschke,
Phys.\ Rev.\ Lett.\ {\bf 102}, 136401 (2009).

\bibitem{Freericks+Krishnamurthy+Sentef+Devereaux}
J.K.~Freericks, H.R.~Krishnamurthy, M.A.~Sentef, and T.P.~Devereaux,
Phys.\ Scr.\ {\bf T165}, 014012 (2015).


\end{thebibliography}
\end{document}